\documentclass[preprint]{elsarticle}
\usepackage{setspace}
\usepackage{graphicx}
\usepackage{natbib}
\usepackage{amsmath, amssymb, mathabx}
\usepackage[colorlinks,urlcolor=blue,citecolor=blue,linkcolor=blue]{hyperref}

\newcommand{\aj}{AJ}
\newcommand{\araa}{ARA\&A}
\newcommand{\mnras}{MNRAS}
\newcommand{\ssr}{Space~Sci.~Rev.}
\newcommand{\nat}{Nature}
\newcommand{\planss}{Planet.~Space~Sci.}
\newcommand{\icarus}{Icarus}
\begin{document}

\begin{frontmatter}
\title{First observations of the Phoebe ring in optical light}
\author[label1]{Daniel Tamayo\corref{cor1}}
\ead{dtamayo@astro.cornell.edu}
\cortext[cor1]{Corresponding Author}
\address[label1]{Department of Astronomy, Cornell University, Ithaca, NY 14853, USA}
\author[label2]{Matthew M. Hedman}
\address[label2]{Department of Physics, University of Idaho, Moscow, ID 83844, USA}
\author[label1]{Joseph A. Burns}

\begin{abstract}
The Phoebe ring, Saturn's largest and faintest ring, lies far beyond the planet's well-known main rings.  It is primarily sourced by collisions with Saturn's largest irregular satellite Phoebe, perhaps through stochastic macroscopic collisions, or through more steady micrometeoroid bombardment.  The ring was discovered with the Spitzer Space Telescope at 24 $\mu$m and has a normal optical depth of $\sim 2 \times 10^{-8}$ \citep{Verbiscer09}.  We report the first observations of sunlight scattered off the Phoebe ring using the Cassini spacecraft's ISS camera at optical wavelengths.  We find that material between $\approx 130-210$ Saturnian radii ($R_S$) from the planet produces an I/F of $1.7 \pm 0.1 \times 10^{-11}$ per $R_S$ of the line-of-sight distance through the disk.  Combining our measurements with the Spitzer infrared data, we can place constraints on the ring-particles' light-scattering properties. Depending on the particles' assumed phase function, the derived single-scattering albedo can match either photometric models of Phoebe's dark regolith or brighter sub-surface material excavated by macroscopic impacts on Phoebe.
\end{abstract}

\end{frontmatter}

\doublespacing

\section{Introduction}
Each giant planet hosts a population of small irregular satellites orbiting at the outskirts of their respective spheres of gravitational influence \citep[see reviews by][]{Jewitt07, Nicholson08}.  These moons are theorized to have been captured early in the Solar System's history when the giant planets were likely migrating \citep{Nesvorny07}, and/or when gas drag was important \citep{Pollack79, Cuk04, Cuk06}.  As a result of being captured, these moons move on high-inclination, high-eccentricity paths that can intersect one another.  This fact, combined with the irregular statellites' anomalously flat size distributions \citep{Bottke10, Kennedy11} suggest a tumultuous collisional history.  Using initial conditions from the Nice model, \cite{Bottke10} find that $99\%$ of the initial mass in the irregular satellites should be ground to dust in the first hundreds of Myr after the moons were captured.

Such circumplanetary disks have important consequences for the larger regular satellites that orbit close to their parent planets.  In the Solar System, the orbits of the circumplanetary dust grains will decay in toward the planet through Poynting-Robertson drag on a timescale of millions of years \citep{Burns79}.  The further-in, and tidally locked, regular satellites then plow through this infalling cloud of dust, modifying their leading hemispheres \citep{Soter74}.  Currently, this is the best explanation for the hemispherical color asymmetries detected on Saturn's moon Iapetus \citep{Denk10, Tosi10, Tamayo11} and on the outer four Uranian regular satellites \citep{Buratti91, Tamayo13a}.  It likely is also ultimately responsible for Iapetus' famous yin-yang albedo pattern \citep{Tamayo11} by triggering a runaway process of ice sublimation \citep{Spencer10}.  Finally, \cite{Bottke13} argue that Jovian irregular-satellite debris explains the dark lag deposits found on the most ancient terrains of Ganymede and Callisto, and that it could be an important source of organic compounds for Europa.  In short, in order to accurately interpret the surfaces of the giant planets' outer main satellites, one must first understand the collisional history of their respective irregular moons.

Despite the present day's drastically reduced collision frequencies between irregular satellites, \cite{Verbiscer09} discovered a vast dust disk around Saturn with the Spitzer Space Telescope.  The height of a collisionally generated disk should correspond to its parent moon's vertical orbital excursions \citep[e.g.,][]{Burns99}, and the disk's height of $\approx 40$ Saturnian radii ($R_S$) implicates the largest irregular satellite Phoebe as the source \citep{Verbiscer09}; however, other smaller satellites that also orbit close to Saturn's orbital plane may also contribute.  This ``Phoebe ring" is extremely diffuse, with a normal optical depth of $\sim 2 \times 10^{-8}$.  Nevertheless, it provides an invaluable opportunity for understanding these circumplanetary debris disks.  The Wide-field Infrared Survey Explorer (WISE) mission has recently obtained a more complete map of the Phoebe ring's emission at a similar wavelength (band 4, centered at 22 $\mu$m) as the 24-$\mu$m band on the Multi-Band Imaging Photometer aboard Spitzer \citep{Skrutskie11}.  However, more measurements at widely spaced wavelengths are needed to constrain the dust grains' properties, such as their wavelength-dependent albedo and emissivity.  \cite{Tamayo12DPS} observed the Phoebe ring with the Herschel Space Observatory at 70 and 130 $\mu$m; unfortunately, due to scattered light from Saturn, we were only able to set upper limits.  In this paper we report the results from our efforts at shorter optical wavelengths.  This is challenging because at these higher energies one measures sunlight scattered by dust into the detector, and one expects dust grains derived from Phoebe to absorb most of the incoming light given the parent moon's low geometric albedo of $\approx 0.085$ across the visible spectrum \citep{Miller11}.  This strongly attenuates an already weak signal.  

We executed the observations with the Imaging Science System's (ISS) Wide-Angle Camera (WAC) aboard the Cassini spacecraft, which has a unique vantage point as it orbits about Saturn.  Relative to observations from Earth, this has the obvious advantage of placing the detector $\sim 300$ times closer to the target.  However, this also implies that from Cassini's location, the full height of the Phoebe ring subtends $\approx 20^{\circ}$ in the sky.  Thus, the debris disk presents a constant background of scattered light across the detector's field of view that cannot be directly measured.  We circumvented this problem by exploiting the shadow cast by Saturn (and its dense rings), which extends behind the planet in a quasi-cylindrical tube.  By capturing the full width of the shadow within a WAC field of view, we measured the scattered light {\it missing} from the region receiving no sunlight, thus indirectly probing the dust content.

\section{Methods} \label{methods}

\subsection{Overview}
\begin{figure}[!ht]
\includegraphics[width=12cm]{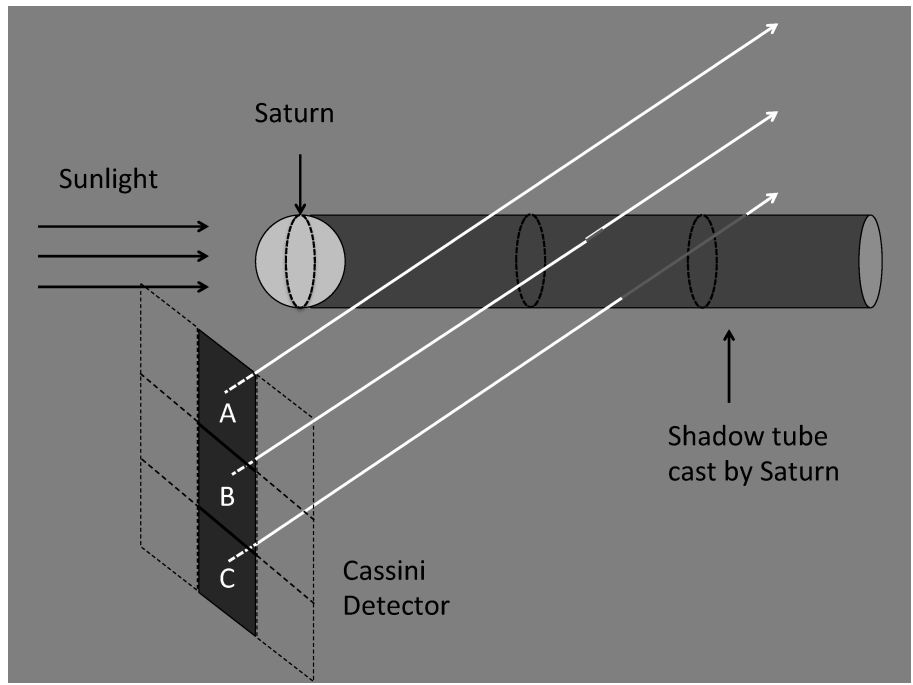}
\caption{\label{cartoon}  The shadow cast by the Saturn system (rings not shown) extends in a quasi-cylindrical tube behind the planet.  Different pixels on the Cassini detector correspond to different lines of sight, shown in white.  As drawn, the line of sight from pixel A misses the shadow tube completely.  B grazes the shadow, so this pixel is only missing the scattered light from a short section of dust and should thus only show a slight brightness decrease relative to A.  C has the longest pathlength through the shadow and should therefore be darkest.  For clarity, the pixel sizes have been exaggerated and the number of pixels has been reduced.  The distances and angles in the diagram are not to scale.}
\end{figure}

As summarized above, we aim to measure the reduction in flux from the Phoebe ring region lying in Saturn's shadow, relative to the background.  The quasi-cylindrical shadow cast by Saturn and its rings pierces the Phoebe ring on the side opposite the Sun, and its full width ($\approx 1^{\circ}$) can be contained in a single WAC field of view ($3.5^{\circ}\times 3.5^{\circ}$).  We acquired 33 220-second WAC exposures using clear filters, i.e., in a band centered at 635 nm \citep{Porco04}.  All images were aimed at the same star field, capturing the section of the shadow from $\approx$ 130 $R_S$ to $\approx$ 300 $R_S$ from Saturn (for details of the data set see Sec. \ref{filter}).  Different pixels in the resulting image represent different lines of sight emanating from the detector that have different pathlengths through the shadow tube (see Fig.\ \ref{cartoon}).  Assuming a constant distribution of dust along the shadow, pixels should register a diminished flux in proportion to their associated pathlengths through the shadow.  This approximation of constant dust density should be valid in the direction perpendicular to the tube's axis, as the shadow's transverse dimensions are much smaller than those of the Phoebe ring.  The magnitude of the radial variation is not well constrained, though the measurements by \cite{Verbiscer09} show the infrared flux is nearly constant on scales of tens of $R_S$, at least in the range 130-180 $R_S$ from the planet (see their Fig.\ 3), so we proceed under this assumption for this initial study.  By measuring the rate at which the flux decreases with increasing pathlength through the shadow, we thus probe the dust content along the tube, together with grain properties like the albedo and phase function.  For details of how we determined the pathlengths through the shadow that correspond to each pixel, see Sec.\ \ref{shadow}.  For an example of an image's modeled pathlengths, and thus of the signature we seek, see the bottom left panel of Fig.\ \ref{sub}.

The sought signal is fainter than that from any ring yet detected in the Solar System.  To motivate our detailed modeling and data analysis below, we first roughly estimate the expected brightness differences between shadowed and non-shadowed regions.  We express all our data as values of I/F, which measures the specific intensity received at the detector relative to the incident solar flux at the Phoebe ring, such that an ideal, diffusely reflective surface would yield an I/F of unity.  

We want to consider the photons that dust particles in the shadow tube {\it would} scatter into the detector were they not in shadow.  Our pathlengths through the tube ($\lesssim 20 R_S$) are comparable to the ring's height ($\approx 40 R_S$), so we take the area filling-factor of dust grains along our line of sight to be roughly the ring's normal optical depth $\tau \sim 10^{-8}$ \citep{Verbiscer09}.  The I/F removed by the shadow is then roughly the product of the particles' albedo and this area filling-factor.  Estimating an albedo $\sim 0.1$ (Phoebe's geometric albedo is $\approx 0.08$, \citealt{Miller11}), yields an I/F $\sim 10^{-10}$.  To put this into perspective, typical I/F values measured from Saturn's extremely faint G-ring (undiscovered until the Voyager flybys) are three orders of magnitude larger than this.  

Standard image processing techniques will fail to extract such a weak signal.  We designed our observations to exploit the fact that, over the $\approx 12$ hours of data collection, the spacecraft's motion causes the shadow to shift position on the field of view by a few tens of pixels, while the stars remain fixed.  By filtering out faulty/noisy pixels (see Sec.\ \ref{filter}) and then subtracting images, we attenuated the much brighter and complex background while retaining a signal from the shifted shadow (see Fig.\ \ref{sub}).  Rather than arbitrarily choosing one of our images as the reference for subtraction, we generated a mean image from our 33 files and subtracted this average field from each of our images.    

\begin{figure}[!ht] \label{sub}
\includegraphics[width=12cm]{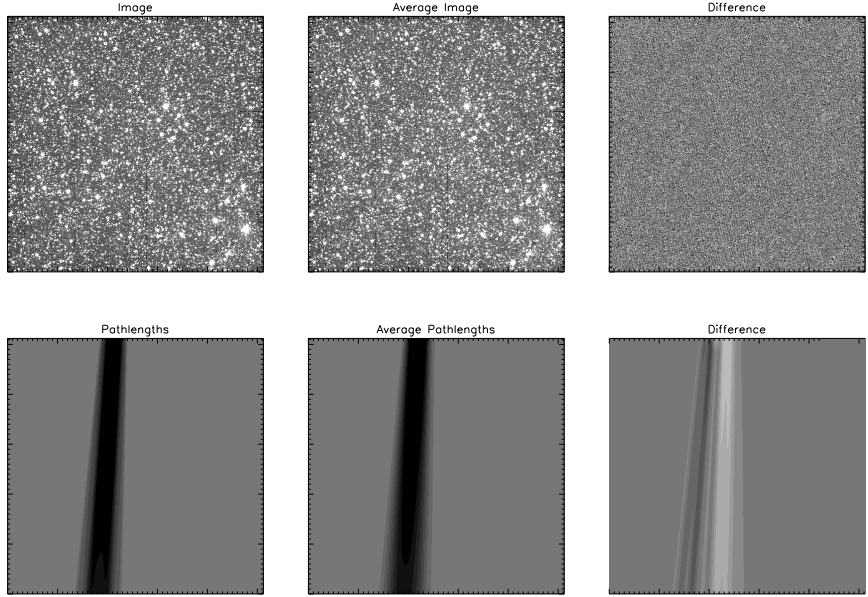}
\caption{\label{sub}  From left to right, the top row of panels shows one of our images, the mean of all 33 images, and the difference between the chosen image and the average (after applying the filtering process described in Sec.\ \ref{filter} and setting flagged pixels to zero).  The grayscale in the first two images represents an I/F range of $[0, 10^{-7}]$, while the difference image is stretched an order of magnitude further, spanning $[-10^{-8}, 10^8]$ .  The subtraction reduces the background level substantially, though a signature is still not discernible by eye.  By zooming in on the top right panel in the electronic version, one can see the filtered pixels in gray, mostly at locations corresponding to stars.  The panels in the bottom row show the modeled pathlengths corresponding to the panels immediately above them (with the color scale inverted to reflect the fact that longer paths through the shadow should appear dark).  The grayscale range for the pathlength plots is $[-30 R_S, 30 R_S]$.  }
\end{figure}

We thus obtain 33 images with the mean field removed, like the one shown in the top right panel of Fig.\ \ref{sub}.  For each image we also calculate the signature expected from the shadow, i.e., the associated pathlength differences for each pixel (inverted since longer pathlengths through the shadow should yield darker pixels---see the bottom left panel of Fig.\ \ref{sub} for an example).   Details of this modeling can be found in Sec.\ \ref{shadow}.  The signature expected in images that have had the mean field removed from them (bottom right panel of Fig.\ \ref{sub}) is then simply the difference between the particular image's pathlength map (bottom left panel) and the mean pathlength map (bottom middle panel).  Pixels with a longer path through the shadow than average should appear darker, while areas traversing less shadow should be brighter.  

Even after applying the above procedure, a signal is not discernible by eye.  However, we can associate a pathlength difference with the I/F measured in each pixel for all 33 images.  We can then bin the $\sim 8$ million pixels according to their associated pathlength differences, and look for a linear trend of decreasing I/F with increasing pathlength difference through the shadow.  The interested reader can skip to Sec.\ \ref{results} for our results.  In the following two sections we provide details of our filtering process and shadow modeling.

We briefly mention that while our method (as we will see) allows us to extract a signal from the Phoebe ring, it imposes an important limitation.  Because the shadow shifts its position between images, a given pixel whose line of sight pierces the shadow 150 $R_S$ from Saturn in one image may cross the shadow at 165 $R_S$ in another.  This means that this pixel in the average image contains information about the shadow over a range of radial distances.  This would confound our method if there were a strong radial gradient in dust concentration, as differences in I/F would no longer be solely determined by differences in path length through the shadow; they could instead reflect variations in the density of dust.  In particular, our method is not well-suited for cases with edges, and WISE images reveal that the Phoebe ring extends to $\sim 270 R_S$ \citep{Hamilton12}.  The radial range in our images varies, but extends from $\approx 130-300 R_S$.  To avoid complicated edge effects, we therefore only used the half of each of our images that pointed closest to the planet in our analysis.  This corresponds to a maximum radial distance from Saturn of $\approx 210 R_S$.  

\subsection{Filtering faulty/noisy pixels} \label{filter}

On day 85 of 2012 (March $25^{\text{th}}$), in Rev 163 (Cassini's $164^\text{th}$ orbit about Saturn), we obtained 33 220-second WAC exposures over $\approx 12$ hours, all aimed at right ascension (RA) $= 210.6^{\circ}$, declination (Dec) $=-7.75^{\circ}$\footnote{Image names W1711398010\_1, W1711399340\_1, W1711400670\_1, W1711402000\_1, W1711403330\_1, W1711404660\_1, W1711405990\_1, W1711407320\_1, W1711408650\_1, W1711409980\_1, W1711411310\_1, W1711412640\_1, W1711413970\_1, W1711415300\_1, W1711416630\_1, W1711417960\_1, W1711419290\_1, W1711420620\_1, W1711421950\_1, W1711423280\_1, W1711424610\_1, W1711425940\_1, W1711427270\_1, W1711428600\_1, W1711429930\_1, W1711431260\_1, W1711432590\_1, W1711433920\_1, W1711435250\_1, W1711436580\_1, W1711437910\_1, W1711439240\_1 and W1711440570\_1.}  This pointing captured a section of Saturn's shadow tube $\approx 130-300 R_S$ from the planet while retaining the same star field across images (see Figs.\ \ref{cartoon} and \ref{sub}).  The data were collected in 2x2 summation mode, yielding images with 512x512 pixels.  As a result, each of our image pixels subtends $1.2 \times 10^{-4}$ rad on a side.  We calibrated our images using the standard Cassini ISS Calibration (CISSCAL) routines \citep{Porco04, West10} to remove instrumental effects, apply flat field corrections and convert the raw data to values of I/F.

Because the sought signal is so faint, it is crucial to pre-process the data to remove noisy pixels and cosmic rays.  After using the CISSCAL calibration routines, we first scanned through each pixel in our 512x512 array and, for a given pixel location, examined the sample of values across our 33 collected images.  We flagged pixels to be discarded if any of the following conditions were met:  (1) the pixel lies on the border of the array, (2) the mean I/F across images was less than 0, (3) any of the 33 values was exactly 0, (4) the mean I/F was above a specified brightness threshold.  Condition (1) was implemented since edge pixels are known to misbehave.  This removed 0.8$\%$ of our pixels.  Condition (2) removes hot pixels that induce errors in the flat-field correction, and disqualified 0.5$\%$ of pixels.  Several horizontal and vertical lines of zero-value pixels could be seen in our images, so we flagged these through condition (3), removing 0.2$\%$ of pixels.  Finally, brighter pixels (in the proximity of stars in the field) have larger dispersions (across the 33 images), in part due to pointing jitter.  We therefore tried a variety of brightness thresholds for condition (4) to optimize the tradeoff between maximizing the number of pixels retained and minimizing the average noise per pixel; we found that a threshold I/F of $6\times10^{-8}$ yielded the lowest $\chi^2$ values in our fits (see Sec.\ \ref{results}).  This reduced the average pixel's standard deviation across images by a factor of 6.4, while flagging $20.2\%$ of pixels.  If we instead chose thresholds of $4\times 10^{-8}$ and $8\times10^{-8}$, this changed our result (the value of the slope we quote in Sec.\ \ref{results}) by less than $3\%$.  

Since we point at a constant RA/Dec, we expect each pixel to exhibit a gaussian distribution about a well-defined mean across our 33 images.  This spatial redundancy can be exploited to remove non-statistical outliers like cosmic rays.  Using only pixels that were not flagged in the first step described in the previous paragraph, we first calculated the standard deviation across our 33 images at each pixel location.  We then removed the largest absolute value from the sample, and recalculated the standard deviation.  If the standard deviation changed by more than $20\%$, we flagged the anomalous pixel, and retained the rest.  We then repeated the process until removing the largest value kept the standard deviation within $20\%$.  The largest number of such iterations required by a pixel in our dataset was 11; however, if a pixel required more than 5 iterations, we flagged the pixel as problematic over all 33 images.  This removed an additional $1.7\%$ of pixels.  Choosing instead standard deviation thresholds of $10\%$ and $30\%$ change our results (the value of the slope we quote in Sec.\ \ref{results}) by less than $4\%$. 

Given our sample of $\sim 8$ million pixels, one would expect no pixels beyond six standard deviations if the distribution was Gaussian.  Therefore, as a final step, we removed pixels with absolute values greater than six times our final distribution's standard deviation.    There were 320 such pixels in our data, which represent $\approx 4\times10^{-3}\%$ of the total.  In total, our combined filtering process removed $23.5\%$ of pixels.

\begin{figure}[!ht]
\includegraphics[width=12cm]{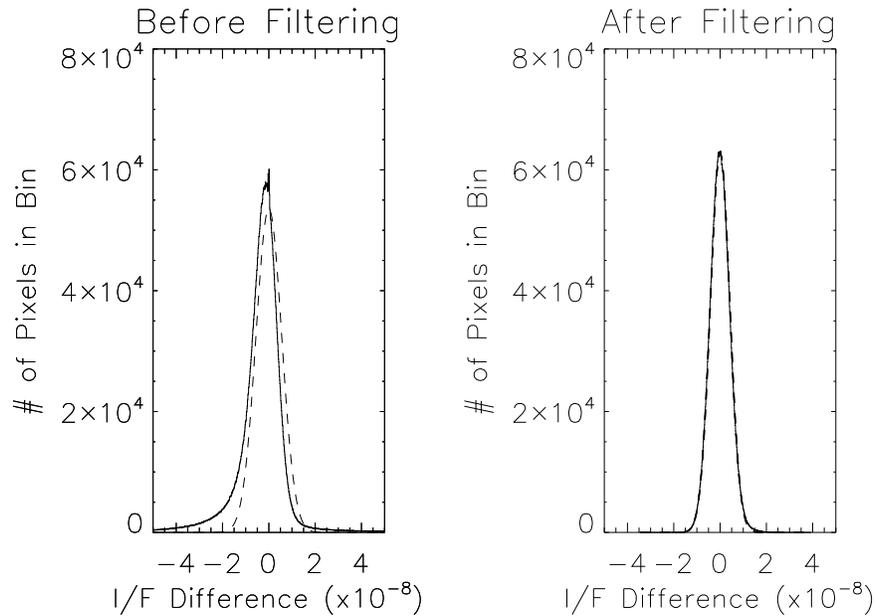}
\caption{\label{stats} Left panel shows the distribution of I/F values in pixels from subtracted images with no filtering applied.  The dashed line shows the best-fit gaussian to the data.  Right panel shows results after filtering.  The best-fit gaussian distribution is also plotted, but is visually indistinguishable from the data.  The bin size is $10^{-10}$.}
\end{figure}

Figure \ref{stats} compares the distribution of brightnesses in our differenced images before and after applying our filtering process.  Both panels include the best-fit gaussian distribution, shown as a dashed line.  The pre-filter histogram (left) exhibits a substantial tail at negative values (removed by our brightness threshold), as well as an overabundance of values close to zero due to faulty pixels.  The accompanying fit also appears skewed due to the presence of outliers at large positive values.  The right panel plots the results after filtering.  We obtain a gaussian distribution about zero that cannot be visually distinguished from the histogram.  The filtered distribution is both tighter and taller because an average image is recomputed after removing problematic pixels that previously skewed the means.  Upon subtracting this refined average field from each of the images, pixels have values closer to zero.

One might worry that as the shadow moves through our 33 images, a pixel which is in shadow in some images but not others might have a high dispersion and be unwittingly flagged by our filtering process.  In fact, the expected lost signal due to the shadow is much smaller (a few times $10^{-10}$) than the standard deviation of the distribution ($4.27\times10^{-9}$), so this is not an issue.  For stronger signals, our procedure would have to be modified.  

\subsection{Modeling the Shadow} \label{shadow}

\begin{figure}[!ht]
\includegraphics[width=12cm]{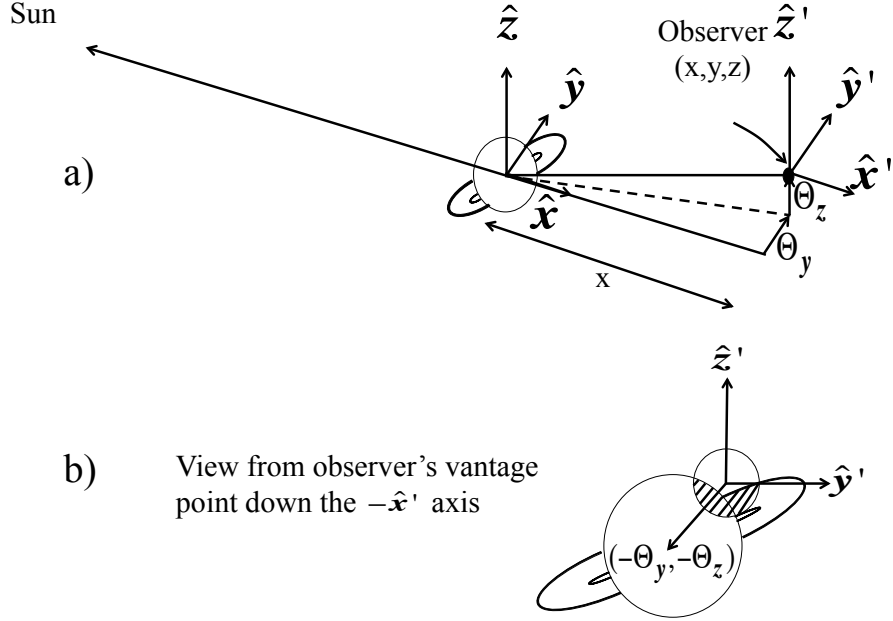}
\caption{\label{geom} a)  The geometry for an observer at position $(x,y,z)$ (or angular coordinates $(\theta_y, \theta_z)$ at distance $x$).  The diagram is not drawn to scale.  We construct a set of axes centered on the observer's position parallel to our coordinate system defined at Saturn (see text).  b)  The view, looking down the $-{\bf \hat{x}'}$ axis, from the observer's location.  The Sun is far enough that the parallax effect is negligible, so it appears centered at the origin.  The observer's displacement from the Sun-Saturn line causes Saturn's apparent angular position to shift by $(-\theta_y, -\theta_z) \approx (-y/x, -z/x)$.  We calculate the shadowing function $S(x,\theta_y, \theta_z)$ by deriving the fraction of the Sun's disk that is occluded by Saturn and its rings at each observer location, e.g., the shaded region in b).}
\end{figure}

We begin by defining our variables and coordinate system.  Let the shadowing function $S$ be the fraction of the Sun's disk that is occluded by Saturn and its rings.  This function will vary from $0$ outside the shadow to $1$ inside the umbra, taking on intermediate values in the penumbra.  We define a coordinate system centered on Saturn where ${\bf \hat{x}}$ points away from the Sun along the Sun-Saturn line, ${\bf \hat{z}}$ lies along Saturn's orbit normal, and ${\bf \hat{y}}$ completes a right-handed triad (Fig.\ \ref{geom}a).  In this frame tied to the Sun-Saturn line, the shadow cast by Saturn and its rings only varies with the Saturnian seasons, owing to the changing cross-section that the planet and annuli present to the Sun's rays.  During our 12-hour observation, the shadow is effectively constant.

We then consider an arbitrary observer with coordinates $(x,y,z)$.  The shadowing function $S(x,y,z)$ is then given by the fraction of the solar disk occluded by Saturn at each position $(x,y,z).$  We calculated $S$ on a $1001\times1001$ pixel grid in the $y-z$ plane spanning 5 $R_S$ in order to capture Saturn's A and B rings.  We lightened the computational load along the $x$ axis by noting that the shadow varies slowly in this direction over the distance range of interest of $120-300 R_S$.  We therefore calculated the shadowing function only every 5 $R_S$ along the $x$ axis, yielding a $1001\times1001\times37$ grid.  For a given $x$ is it easiest to calculate the shadowing function in angular space, yielding $S(x, \theta_y, \theta_z)$, where $\theta_y \approx y/x$ and $\theta_z \approx z/x$ (see Fig.\ \ref{geom}a).  

As the observer moves to different values of $(\theta_y, \theta_z)$, the angular position of Saturn will vary relative to the fixed stars.  By contrast, the Sun is far enough away that the parallax effect is negligible given our effective pixel size of $1.2 \times 10^{-4}$ rad.  We therefore define a local coordinate system $(\hat{x}', \hat{y}', \hat{z}')$ at $(x,y,z)$ with reference axes parallel to our previous ones centered on Saturn (see Fig.\ \ref{geom}a).  For all observers that we consider, then, the Sun lies along the $-\hat{x}'$ direction.  We could construct a spherical coordinate system with the pole along the $-x'$ axis, but because $\theta_y$ and $\theta_z$ are small, we can approximate the problem in flat space, treating $\theta_y$ and $\theta_z$ as linear distances in the $y-z$ plane.  Saturn, then, will be centered at $(-\theta_y, -\theta_z)$, see Fig.\ \ref{geom}b.

To obtain $S(x,\theta_y, \theta_z)$, we then calculate the fraction of the Sun's disk that is blocked by Saturn.  We define a function $\odot(\theta_y',\theta_z')$ with value unity inside the solar disk (assumed uniform and circular) and 0 outside, and a similar function $\Saturn(x, \theta_y',\theta_z')$ with value unity inside the cross-section of Saturn and its rings.  As opposed to the Sun's angular size, which does not change appreciably as we vary $x$ over the range of interest, Saturn's angular size will change substantially, so $\Saturn$ is a function of $x$.  To calculate $S(x, \theta_y,\theta_z)$, we  then simply offset $\Saturn(x, \theta_y',\theta_z')$ by $(-\theta_y, -\theta_z)$, point-wise multiply $\odot$ and $\Saturn$, and integrate over all of $(\theta_y', \theta_z')$ space (See Fig.\ \ref{geom}b).  Recalling that all angles are being approximated as linear distances in flat space, and expressing $S$ as the fraction of the solar disk that is occluded,
\begin{equation}
S(x, \theta_y,\theta_z) = \frac{1}{{A_\odot}} \int{\odot(\theta_y',\theta_z')\Saturn(\theta_y' + \theta_y, \theta_z' + \theta_z)d\theta_y'd\theta_z'},
\end{equation}
where ${A_\odot}$ is the area of the Sun's disk, or $\pi{\theta_\odot}^2$, with $\theta_\odot$ the Sun's angular size $\approx 10^{-3}$ rad.  

The computational method matches one's intuition that the umbra should narrow and the penumbra widen with increasing $x$.  As the observer moves away from Saturn, the angular function $\Saturn$ scales as $x^{-1}$.  Thus, as $x$ increases, Saturn's angular size relative to the Sun shrinks, and Saturn does not completely block the solar disk over a larger area of $(\theta_y, \theta_z)$ space, i.e., the penumbra becomes larger.  

Because our Sun model is symmetric, i.e., $\odot(\theta_y',\theta_z') = \odot(-\theta_y',-\theta_z')$, we can write the shadowing function as a convolution by letting $\theta_y' = -\theta_y'$ and $\theta_z' = -\theta_z'$:
\begin{eqnarray}
S(x, \theta_y, \theta_z) &=& \frac{1}{{A_\odot}} \int{\odot(-\theta_y',-\theta_z')\Saturn(\theta_y - \theta_y', \theta_z - \theta_z')d\theta_y'd\theta_z'} \\
&=& \frac{1}{{A_\odot}} \int{\odot(\theta_y',\theta_z')\Saturn(\theta_y - \theta_y', \theta_z - \theta_z')d\theta_y'd\theta_z'} \\
&=& \frac{1}{\pi {\theta_\odot}^2} \odot \mathop{*} \Saturn.
\end{eqnarray}

We thus have a compact way of calculating $S$ that can be immediately converted to linear coordinates through $y=x\theta_y$ and $z=x\theta_z$.  For the angular function $\Saturn(x, \theta_y',\theta_z')$, we modeled the planet as an oblate spheroid, and included the A and B rings, assuming them to be perfectly opaque.  We calculated the appropriate cross-section of our Saturn-system model perpendicular to the Sun-Saturn line using the Navigation and Ancillary Information Facility (NAIF) SPICE toolkit \citep{Acton96}.

We now briefly estimate the error on $S$.  If the Sun were a point source, $S$ would always be 0 or 1.  Thus, the error on $S$ in the penumbra is fundamentally set by our pixel size's ability to resolve the solar disk.  The solar disk is spanned by $\approx 10$ pixels, so the error in $S$ is $\sim 10\%$.  This base limitation allows us to ignore several complications not discussed above.  We found that the following effects are unimportant at the $10\%$ level:  aberration of light, light travel time, atmospheric deviations in Saturn's shape from an oblate spheroid, the slight variations in the Saturn system's cross-section as viewed from different positions in our grid $(x,y,z)$, and the fact that the A ring is slightly transmissive at the relevant incidence angle.

With our 3-dimensional model in hand, we then proceeded to calculate the pathlengths through the shadow for the various lines of sight corresponding to each of the pixels in our images.  We first geometrically navigated the images using the stars in the field, and then calculated the RA/Dec coordinates for each of our pixels.  Finally, for each pixel, we numerically integrated $S$ along the line of sight to yield the associated pathlength through the shadow, generating 512x512 arrays like the ones displayed on the bottom row of Fig.\ \ref{sub}.  
  
\section{Results} \label{results}

\begin{figure}[!ht]
\includegraphics[width=12cm]{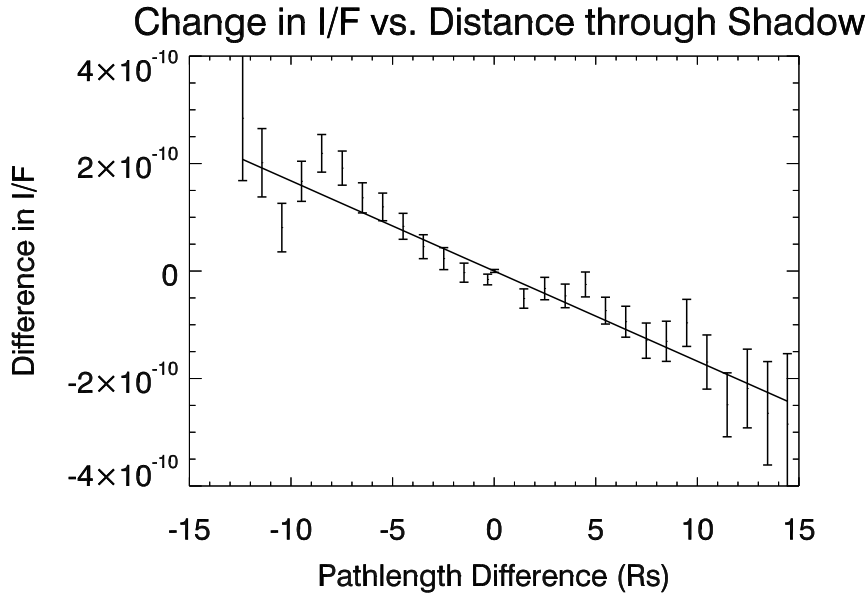}
\caption{\label{linear} Differences in I/F vs. differences in pathlength through the shadow.  As expected, pixels that see through more of the shadow than average have lower I/F values than pixels that see through less.  The best-fit slope of a linear fit is $m = -1.7 \pm 0.1 \times10^{-11}/R_S$.  The reduced $\chi^2$ is 1.38 with 26 degrees of freedom.  We attribute this high $\chi^2$ value to radial gradients in dust concentration in the Phoebe ring.}
\end{figure}

After the pre-processing described above, we bin our sample of $\approx 3.3$ million surviving pixels according to their associated pathlengths through the shadow.  Fig.\ \ref{linear} plots the mean value in each bin, with their associated standard errors.  As can be surmised from the bottom panels of Fig. \ref{sub}, most pixels ($\approx 2.5$ million) correspond to lines of sight that do not pierce the shadow.  This explains the extremely tight error bar on the zero bin.  The bins from $\approx$ -10 to 10 $R_S$ have $\sim 10^4$ values, thus improving on the error per pixel by a factor of $\sim$ Sqrt($10^4$), and explaining why a clear signal is seen in the binned data despite the signature not being discernible in the images (Fig. \ref{sub}).

As expected, there is a definite trend toward lower I/F values with increasing pathlength through the shadow.  The best-fit slope is $m = -1.7 \pm 0.1 \times10^{-11}/R_S$.  The absolute value of this rate can be more straightforwardly interpreted as the I/F generated by dust grains in the Phoebe ring per $R_S$ of the line-of-sight distance through the disk. 

The reduced $\chi^2$ from the fit is 1.38, with 26 degrees of freedom, so to be conservative we quote our statistical error on the slope multiplied by a factor of the square root of the reduced $\chi^2$.  We attribute this high $\chi^2$ to radial variations in dust concentration along the shadow tube, which our simple model assumes do not exist.  This is a difficult problem to disentangle.  Since Cassini resides close to the planet ($\sim 20 R_S$) and the Phoebe ring is far ($\sim 200 R_S$), we are looking nearly down the axis of the shadow.  While this helps to extract the exceedingly faint signal by providing longer lines of sight through the shadow, it also causes the loss of radial information since each pixel samples a column of dust over a range of distances from Saturn $\sim 20 R_S$ long.  Furthermore, as discussed in the previous section, by subtracting images from one another to attenuate the bright background, we sometimes compare measurements from radii varying by as much as $40 R_S$.  

\begin{figure}[!ht]
\includegraphics[width=12cm]{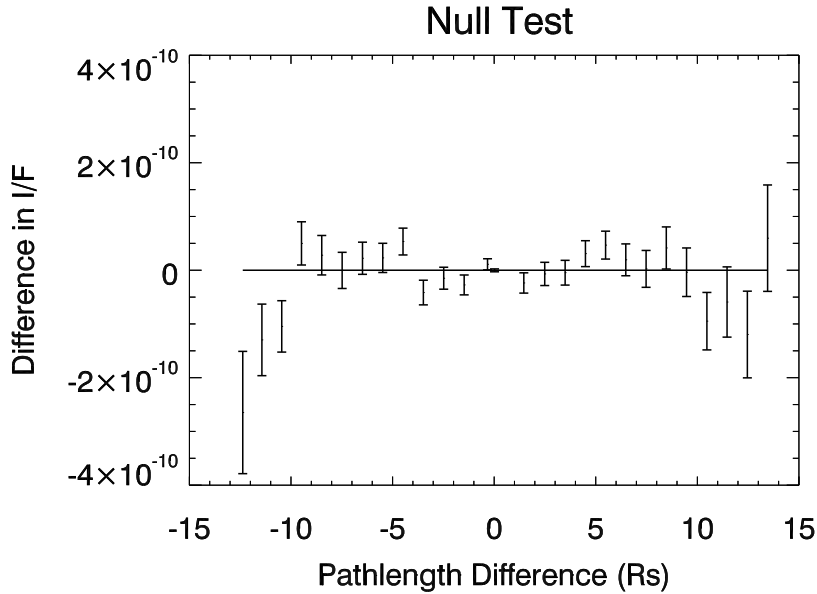}
\caption{\label{null}  Null test performing the same analysis that led to Fig.\ \ref{linear}, but with the shadow model offset by 200 pixels.  As expected, there is no correlation in I/F deviations with pathlength differences through the incorrectly placed shadow model.  A constant-value hypothesis provides the best reduced $\chi^2$, and a linear fit gives a slope consistent with zero.}
\end{figure}

To validate the robustness of our result, we perform the same analysis, but with the shadow model (see Fig.\ \ref{sub}) offset to the right by 200 pixels---where, of course, there is no shadow.  In this case we expect to see no correlation between our modeled pathlengths and the measured deviations in I/F, as all the dust in this section of the field of view is in full sunlight.  Figure \ref{null} shows the analogous plot to Fig.\ \ref{linear}.  The reduced $\chi^2$ for a constant-value model is 1.68 with 27 degrees of freedom, and a linear fit gives a slope consistent with zero (and a slightly higher reduced $\chi^2$).  We again attribute the correlated variations to radial gradients in the dust concentration that similarly affect our method in this part of the ring under full sunlight.

Since our shadow grid is only determined to within a pixel, we estimate the systematic uncertainty in our modeling by offsetting the calculated pathlengths (bottom left panel of Fig.\ \ref{sub}) by one pixel (along each of eight possible directions) and recalculating the slope from Fig.\ \ref{linear}.  The mean slope across these analyses is $m = -1.7 \times 10^{-11}/R_S$, identical to the value found above.  The standard deviation is less than $1\%$, implying that pointing and modeling uncertainties do not dominate our errors.

\subsection{Photometry} \label{phot}

In order to estimate the disk's optical depth, \cite{Verbiscer09} had to make assumptions about the Phoebe-ring grains' particle-size distribution, albedos, and emissivities.  Our new measurement at optical wavelengths still leaves the problem underdetermined, but we can combine the optical and infrared data to place rough constraints on the particles'  light-scattering properties.  Since we only observe sunlight scattered almost directly backward at a Sun-ring-observer angle, or phase angle, of $\alpha \approx 6^{\circ}$, we ignore the diffracted component of light in our subsequent analysis.  Furthermore, because particles with radii $s \lesssim 5 \mu$m (assuming spherical grains) are quickly removed from the Phoebe ring by radiation pressure \citep{Tamayo11}, particles are much larger than the optical wavelengths ($\lambda$) at which we observe, so we are in the geometric optics limit.  Thus, the single-scattering albedo that we determine assumes a geometric cross-section $\sigma = \pi s^2$ to calculate the power incident on dust grains, and, when accounting for the outgoing power, ignores the light that is diffracted forward (at high phase) into a cone of angular width $\approx \lambda/(2s)$.  

For low optical-depth clouds, the measured I/F is related to the line-of-sight optical depth $\tau$ and the single-scattering albedo $\varpi_0$ (at 0.635 $\mu$m, where our band is centered) through \begin{equation} \label{IF}
\frac{I}{F} = \frac{1}{4} \tau \varpi_0 P(\alpha),
\end{equation}
where $P(\alpha)$ is the phase function \citep[e.g.,][]{Burns01}.  Since we are in the geometric optics limit, $\tau$ is a geometrical optical depth and, for low $\tau$, is approximately the area filling-factor of dust along the line of sight.  In our observations we obtain the differential change in I/F with distance, so we write $d\tau = \eta \: dl$, where $dl$ is a differential length element along the line of sight and $\eta$ is given by 
\begin{equation}
\eta = \int{n \sigma ds},
\end{equation}
with $n\:ds$ the number density of dust grains with physical radii between $s$ and $s+ds$ and $\sigma$ the geometrical cross-section $\pi s^2$.  We thus obtain
\begin{equation} \label{m}
m \equiv \frac{d(I/F)}{dl} = \frac{\eta\varpi_0 P(\alpha) }{4},
\end{equation}
where $m$ is our measured slope of $m = -1.7 \pm 0.1 \times10^{-11}/R_S$.

In the case of the Spitzer observations at 24 $\mu$m, where one observes the grains' thermal emission, the particle sizes become important.  A spherical blackbody in the Phoebe ring would have an equilibrium temperature of $\approx 90$K and emit $\approx 90\%$ of its energy in the wavelength range $\lambda = 10-100 \mu$m.  For typical particle-size distributions, one mostly observes the smallest grains (in this case $s \sim 5 \mu$m) that dominate the population's surface area.  Thus, the dimensionless parameter $X = 2\pi s / \lambda \sim 1$ over the wavelength range in which blackbody grains would preferentially emit, so real particles will have difficulty releasing energy at these long wavelengths.  These small grains must therefore heat up beyond their equilibrium blackbody temperatures in order to release the energy they absorb.  

Given our limited data, we follow the simple model of \cite{Verbiscer09} using a constant infrared emissivity $\epsilon$.  Energy balance then requires
\begin{equation} \label{energy}
\pi F \pi s^2 (1-A) = 4\pi s^2 \epsilon\sigma_B T^4,
\end{equation}
where $\pi F$ is the solar flux at Saturn, $\sigma_B$ is the Stefan-Boltzmann constant, T is the equilibrium grain temperature, and $A$ is the bolometric Bond albedo, which is integrated over all phase angles and wavelengths, and weighted by the solar spectrum.  Since the solar spectrum peaks in the optical, and Phoebe's geometric albedo is flat across the visible spectrum \citep{Miller11}, we can reasonably approximate the bolometric Bond albedo by the Bond albedo at 635 nm, where our observing band is centered.  We can then relate $A$ to the single-scattering albedo $\varpi_0$ through $A = \varpi_0 P(0) q / 4$, where $q$ is the phase integral.  Rearranging Eq. \ref{energy}, 
\begin{equation} \label{T}
T = \Bigg(\frac{\pi F(1-A)}{4\epsilon \sigma_B}\Bigg)^{\frac{1}{4}}.
\end{equation}
The corresponding emission at Spitzer's 24 $\mu$m band is $\epsilon B_\nu(24 \mu\text{m}, T)$, where $B_\nu$ is the Planck function.  The ratio of the observed intensity with Spitzer to $\epsilon B_\nu$ then gives the area filling-factor of dust grains along the line of sight, or the geometrical optical depth $\tau$.  If we assume that the number density of dust grains does not vary within the Phoebe ring (the same assumption used in our above analysis), then $\tau = \eta \: L$, where L is the total pathlength through the disk.  Given the Phoebe ring extends to $\sim 270 R_S$ \citep{Hamilton12}, the Spitzer observations that pierce the disk edge-on at $\approx 150 R_S$ imply $L \sim 400 R_S$.  Combining the above relations we obtain
\begin{equation}
\eta = \frac{I_{Sp}}{\epsilon B_\nu(T) L},
\end{equation}
where $I_{Sp}$ is the intensity measured by Spitzer, and $T$ is given by Eq.\ \ref{T}.  Finally, plugging $\eta$ into Eq.\ \ref{m} and rearranging we have
\begin{equation}
4 \epsilon mL \frac{B_\nu(T)}{I_{Sp} P(\alpha)} = \varpi_0,
\end{equation}
where $T$ depends on $\varpi_0$ implicitly through Eq.\ \ref{T}.  

If we treat Phoebe ring particles as isotropic scatterers ($P(\alpha) = 1$), the above relations are simplified, since in this case $A = \varpi_0$.  If like \cite{Verbiscer09} we then assume $\epsilon = 0.8$, we obtain $\varpi_0 \approx 0.2$.  If instead we vary $\epsilon$ from 0.1-1, and $L$ from $350-450R_S$, $\varpi_0$ ranges from $\approx$ 0.2 (high emissivity, low pathlength) to $\approx$ 0.3 (low emissivity, high pathlength).  This range is higher than the corresponding values inferred for Phoebe regolith particles from photometric modeling of Phoebe observations at 0.48 $\mu$m \citep{Simonelli99} and 0.9-1.4 $\mu$m \citep{Buratti08}, which both yielded $\varpi_0 \approx$ 0.07.

Several possibilities could account for this discrepancy.  The assumption that particles are roughly isotropic scatterers is plausible.  If one ignores the diffracted component of light (which causes the phase functions of small grains to rise at high phase angles), \cite{Pollack80} found that irregularly shaped dust grains have approximately flat phase functions, and empirical fits to the phase function of particles in Saturn's faint G-ring also yield roughly isotropic scatterers (M.M. Hedman, 2013, private communication).  If this accurately represents Phoebe ring particles, our high single-scattering albedo may point to the impacts that generated the disk having excavated brighter, sub-surface material.  Indeed, high-resolution images taken during the Phoebe flyby as Cassini approached Saturn reveal bright material lining crater walls \citep{Porco05}.  \cite{Buratti08} report I/F values at 0.9 $\mu$m 4-5 times larger for the bright material relative to the predominantly dark surface.  Further photometric modeling is required to quantitatively compare the bright material's surface reflectance with the single-scattering albedo of the regolith particles that make it up.  

Alternatively, if the dust's phase function were not flat, but instead rose at small phase angles, our assumption above of isotropic particles would cause us to infer an artificially high single-scattering albedo, as our measurement at low phase would represent more than its fair share of the total scattered light.  For example, if we assume a Henyey-Greenstein function \citep{Henyey41} with parameter g = -0.35 (back-scattering), which in a Hapke model can reproduce Phoebe's photometry \citep{Miller11}, we find $\varpi_0 \approx 0.06$ (taking the above nominal values of $\epsilon = 0.8$ and $L = $ 400 $R_S$).  This is consistent with the modeled regolith albedo of $\varpi_0 \approx 0.07$ found by \cite{Simonelli99} and \cite{Buratti08}.

Another possibility is that photometric model fits like those used in the above-quoted studies to obtain $\varpi_0 \approx 0.07$ do not accurately reproduce the light-scattering properties of the grains that form the Phoebe ring.  Laboratory experiments by \cite{Shepard07} show that Hapke models cannot reliably be used to uniquely infer regolith properties; however, they found that the single-scattering albedo of regolith particles is the most robust parameter extracted.  It is therefore unlikely that this could fully account for the difference.

Finally, it is possible that smaller, brighter, irregular satellites also contribute to the Phoebe ring.  The only observable currently connecting the disk to Phoebe is its vertical extent, which matches Phoebe's vertical excursions on its orbit.  However, the moons Ymir, Suttungr, Thrymr, and Greip all have comparable orbital inclinations to Phoebe and are candidate sources.  However, the albedos of irregular satellites that have been measured are low \citep{Grav13}, so, while these moons may nevertheless contribute to the ``Phoebe" ring, they seem unlikely to substantially raise its albedo.

\section{Conclusion}

We have described the first measurements of the Phoebe ring in optical light.  Extracting the exceedingly faint signal (I/F variations $\sim 10^{-10}$) was only possible by subtracting multiple images of the same star field, thereby attenuating the relatively bright background while retaining a signature from Saturn's shifting shadow.  A careful statistical analysis then allowed us to indirectly measure the I/F generated by scattering Phoebe-Ring dust grains per unit pathlength through the disk.  We obtained a value of $m = 1.7 \pm 0.1 \times 10^{-11}R_S$; thus, for example, a line of sight 100 $R_S$ long would generate an I/F of $\approx 1.7 \times 10^{-9}$.

In Sec.\ \ref{phot} we then combined our measurement with the infrared intensity measured with Spitzer \citep{Verbiscer09} to constrain the grain albedos.  Assuming particles to be isotropic scatterers, we derive albedo values higher than those obtained from photometric models applied to observations of Phoebe \citep{Simonelli99, Buratti08}.  This may suggest that the impact(s) that generated the Phoebe ring are excavating bright sub-surface material, as observed on some of the moon's crater walls \citep{Porco05}.  Alternatively, our measurements can be brought into agreement if the phase function of dark Phoebe ring grains follows a Henyey-Greenstein function with parameter g = -0.35, which \cite{Miller11} used to match the photometry of Phoebe using a Hapke model.  The former hypothesis may implicate one or several larger collisions in the formation of the Phoebe ring, in order to effectively sample the sub-surface.  The latter is consistent with a roughly steady state of micrometeoroid bombardment.

Our measurements spanned the range $\approx 130-210 R_S$.  We plan to make future measurements of the Phoebe ring closer to Saturn with Cassini, and taking a larger number of exposures.  The increased signal-to-noise should allow us to generate a radial profile of the disk, which may allow us to differentiate between a steady-state model of micrometeoroid bombardment and one invoking large, stochastic impacts.  This would inform our interpretation of the grain properties discussed above.  

We also plan to observe both inside and outside the orbital path of the two-faced moon Iapetus (at $59 R_S$), to verify whether the satellite sweeps up most of the infalling debris \citep{Tamayo11}.  This would observationally settle a puzzle that has existed since the moon's discovery over 300 years ago.  It is worth noting that this measurement can presently only be made by the Cassini spacecraft, given its favorable position about Saturn.  The sought signal is so faint that scattered light from the planet should preclude currently planned Earth-bound telescopes from observing it inside $\sim 60-80 R_S$.   

\section{Acknowledgements}
We would like to thank Philip D. Nicholson, Michael W. Evans, Matthew S. Tiscareno, and Rebecca A. Harbison for helpful and insightful discussions.  We would also like to thank the two anonymous reviewers who greatly strengthened and clarified the final manuscript.  We gratefully acknowledge support from the Cassini mission.

\end{document}